\documentclass[reprint,aps,prc,twocolumn,superscriptaddress,nolongbibliography,floatfix,10pt]{revtex4-2}
\usepackage{xeCJK}
\usepackage{epsfig,amsmath,graphicx,amssymb,tabularx}
\usepackage[bookmarks=true,colorlinks,citecolor=blue,linkcolor=blue,anchorcolor=blue,filecolor=blue,urlcolor=blue]{hyperref}
\usepackage[T1]{fontenc}
\usepackage[utf8]{inputenc}
\usepackage{dcolumn}
\usepackage{bm}
\usepackage{multirow}
\makeatletter
\def\NAT@def@citea{\def\@citea{\NAT@separator}}
\makeatother

\begin{document}
\title{Microscopic quasifission dynamics of the ${}^{54}\text{Cr}+{}^{243}\text{Am}$ reaction}
\author{Liang Li (李良)}
\affiliation{School of Nuclear Science and Technology, University of Chinese Academy of Sciences, Beijing 100049, China}
\author{Lu Guo (郭璐)}
\email{luguo@ucas.ac.cn}
\affiliation{School of Nuclear Science and Technology, University of Chinese Academy of Sciences, Beijing 100049, China}
\date{\today}
\begin{abstract}
\edef\oldrightskip{\the\rightskip}
\begin{description}
\rightskip\oldrightskip\relax
\setlength{\parskip}{0pt} 

	\item[Background] The synthesis of superheavy elements (SHEs) beyond Oganesson, such as $Z=119$, remains a formidable challenge, primarily because the dominant quasifission (QF) channel severely hinders the formation of compound nuclei. A microscopic understanding of the QF dynamics is therefore essential for guiding future experiments.
		
	\item[Purpose] We investigate the microscopic mechanisms of QF in the ${}^{54}\text{Cr}+{}^{243}\text{Am}$ reaction, a key candidate system for synthesizing SHE 119, with particular emphasis on the roles of projectile orientation and incident energy.
		
	\item[Method] The calculations are performed using the fully microscopic time-dependent Hartree-Fock theory based on the Skyrme energy density functional. We perform systematic simulations covering a broad set of initial orientations of the deformed ${}^{54}\text{Cr}$ and ${}^{243}\text{Am}$ nuclei, together with a finely spaced range of incident energies extending from below to well above the Coulomb barrier.
	
\item[Results] Our fixed-energy simulations show that projectile side collisions are strongly shell driven, with heavy and light fragments tending toward the spherical $Z=82$ and deformed $N=52\text{--}56$ shell regions, respectively, whereas tip collisions exhibit weaker shell influence. These shell-driven reactions are associated with shorter contact times, consistent with faster neck rupture caused by the enhanced rigidity of shell-stabilized fragments. In the tip-tip channel, the energy dependence further reveals a transition from an octupole-deformed region around $Z \approx 88$ to a spherical shell-driven regime at higher energy, while shell steering is comparatively weak in a near-barrier region at a center-of-mass energy of approximately 226~MeV.

 \item[Conclusions] Our microscopic study demonstrates that the manifestation of shell effects in the final QF fragment distributions is a dynamical outcome sensitively dependent on both the initial collision geometry and the incident energy. 
 This pronounced energy dependence indicates that the competition between QF and fusion may be affected by the beam energy. 
 The comparatively weak shell steering found in the near-barrier tip-tip region therefore suggests a possible low-energy window of potential interest for SHE synthesis.

\end{description}
\end{abstract}
\maketitle
\section{INTRODUCTION}
\label{intro}
The synthesis of superheavy elements (SHEs) has long been a central pursuit in nuclear physics, motivated by the quest to determine the upper limit of atomic number and to verify the existence of the ``island of stability'' governed by quantum shell effects~\cite{Sobiczewski2007,smits2024}. To date, SHEs up to $Z = 118$ have been successfully synthesized through heavy-ion fusion reactions, using either ``cold fusion'' with $^{208}$Pb or $^{209}$Bi targets or ``hot fusion'' with a $^{48}$Ca beam on actinide targets~\cite{hofmann2000,morita2004,oganessian2006,huang2026}.
Despite these achievements, all major attempts over the past decade to extend the periodic table into the eighth period by synthesizing elements 119 and 120 have been unsuccessful~\cite{oganessian2009,hofmann2016,khuyagbaatar2020,tanaka2022}. This challenge primarily stems from the fact that $^{48}$Ca-induced reactions are no longer feasible due to the limited availability of target isotopes such as einsteinium and fermium. As a result, heavier projectiles, such as $^{50}$Ti, $^{51}$V, and $^{54}$Cr, are now being employed in reactions with actinide targets.
Recently, the Lawrence Berkeley National Laboratory (LBNL) in the United States successfully synthesized $^{290}$Lv via the $^{50}$Ti + $^{244}$Pu reaction, demonstrating the feasibility of using projectiles heavier than $^{48}$Ca for SHE synthesis~\cite{gates2024}. This breakthrough has renewed confidence in the pursuit of elements 119 and 120 using similar reactions. Meanwhile, the Institute of Modern Physics (IMP), Chinese Academy of Sciences is preparing to explore element 119 through the $^{54}$Cr + $^{243}$Am system at the upgraded HIRFL-CAFE2 facility~\cite{gan2022}.

While these experimental breakthroughs have sparked renewed optimism, synthesizing elements 119 and 120 remains an exceptionally challenging task. The primary obstacle stems from the combined prohibitive effects of quasifission (QF) and fusion--fission. QF dominates the dynamical evolution from the captured dinuclear system toward a fully equilibrated compound nucleus. This causes the system to re-separate before achieving full equilibration, thereby significantly suppressing the formation probability~\cite{godbey2020,hinde2021,sun2023b}. Even if a compound nucleus is formed, its low fission barrier makes it highly susceptible to fission. As a result, evaporation-residue cross sections are suppressed to the femtobarn level (or even lower) for the synthesis of elements 119 and 120. 
Providing reliable theoretical predictions for these reactions presents a formidable challenge, as QF dynamics arise from a subtle interplay between entrance- and exit-channel degrees of freedom, including the collision energy~\cite{back1983,oberacker2014,sun2022b}, the deformation and orientation of the colliding nuclei~\cite{hinde1995,nishio2012,wakhle2014}, the neutron-to-proton asymmetry of the composite system~\cite{umar2017,guo2018c,simenel2020,sen2022}, and the shell effects in the emerging fragments~\cite{morjean2017,mohanto2018,li2022}. These factors, individually or collectively, govern the trajectory of the dinuclear system and determine whether it evolves toward fusion or re-separates.

To capture these intricate dynamics, the theoretical description of QF spans a wide spectrum of models. On the macroscopic side, approaches such as the dinuclear system model~\cite{adamian2003,wang2012,zhu2022,deng2023,wang2026}, stochastic Langevin-type equations~\cite{zagrebaev2007}, and the dynamical cluster-decay model~\cite{chopra2021} have been successfully developed. In these phenomenological frameworks, the complex many-body problem is effectively reduced to the evolution of a few key collective degrees of freedom, such as the internuclear distance and mass asymmetry, driven by conservative potentials and dissipative forces. On the microscopic front, semi-classical transport theories, including the Quantum Molecular Dynamics~\cite{wang2016,chen2024,feng2025} and Boltzmann-Uehling-Uhlenbeck models~\cite{feng2023}, are widely employed to simulate the evolution based on the motion of nucleons and explicit nucleon-nucleon collision terms. By tracking the phase-space evolution and incorporating two-body scattering, these transport approaches provide robust descriptions of fragmentation and fluctuation phenomena. Furthermore, advanced frameworks, including stochastic mean-field theory~\cite{ayik2021,sekizawa2025} and the time-dependent random phase approximation~\cite{balian1984,williams2018} which extend beyond the standard mean-field approximation, as well as relativistic covariant density functional theory~\cite{zhang2024}, have also been applied to QF studies.

Among these approaches, the time-dependent Hartree-Fock (TDHF) theory stands out as the most foundational microscopic framework for low-energy heavy-ion dynamics~\cite{nakatsukasa2016,simenel2018,stevenson2019,sekizawa2019,sun2022c,simenel2025a}. In contrast to the macroscopic and semi-classical models discussed above, TDHF provides a fully microscopic, self-consistent description without the need for empirical parameters specific to the reaction mechanism~\cite{negele1982,guo2007,simenel2007,guo2008,simenel2008}. By dynamically evolving the self-consistent mean field, the theory naturally incorporates essential physical effects such as nuclear deformation and orientation dependence~\cite{umar2015a,sekizawa2016,stevenson2022,simenel2025c,wu2026}, nucleon transfer~\cite{sekizawa2013,wu2019,li2019,jiang2020,simenel2025b}, and one-body dissipation~\cite{simenel2010,dai2014,sekizawa2019b}. Consequently, TDHF has achieved remarkable success in reproducing experimental observables across a wide range of phenomena, including fusion~\cite{stevenson2016,godbey2019c,sun2022,sun2023,yao2024,jiang2025b}, fission~\cite{simenel2014a,goddard2015,huang2024,huang2024b,qiang2025}, multinucleon transfer~\cite{sekizawa2017a,wu2020,wu2022}, resonance dynamics~\cite{simenel2001,reinhard2007,simenel2009} and, crucially, QF reactions~\cite{umar2016,guo2018c,simenel2021}. Recent TDHF studies have demonstrated close agreement with data on mass--angle correlations and fragment mass distributions~\cite{hammerton2015,umar2015c,yu2017}, while offering deep insights into the steering role of spherical and deformed shell closures~\cite{godbey2019,li2024c,simenel2024}. These successes firmly establish TDHF as a powerful predictive tool to investigate the microscopic mechanisms governing the QF process.

Motivated by these successes, in this work, we apply the TDHF method to the $^{54}$Cr + $^{243}$Am system, a prime candidate for synthesizing element 119. Our objective is to provide a comprehensive microscopic investigation of the QF mechanisms in this reaction, with a specific focus on the combined influence of the incident energy and the initial orientations of the deformed reactants. The remainder of this article is organized as follows: Section~\ref{theory} outlines the theoretical framework and computational details. The simulation results and detailed discussions are presented in Section~\ref{results}. Finally, a summary of our findings is provided in Section~\ref{summary}.
\section{THEORETICAL FRAMEWORK}
\label{theory}
The fundamental assumption of the TDHF theory is that the nuclear many-body wave function $\Psi(t)$ remains a single Slater determinant at all times,
\begin{equation}
\Psi(t) = \mathcal{A} \left( \phi_1(\mathbf{r}_1, t) \phi_2(\mathbf{r}_2, t) \dots \phi_A(\mathbf{r}_A, t) \right),
\end{equation}
where $\mathcal{A}$ is the antisymmetrization operator and $\phi_i(\mathbf{r}, t)$ are the $A$ occupied single-particle wave functions. The time evolution of these wave functions is governed by
the TDHF equations,
\begin{equation}
	i\hbar \frac{\partial}{\partial t} \phi_i(\mathbf{r}, t) = \hat{h}(t) \phi_i(\mathbf{r}, t),
\end{equation}
which can be derived from the time-dependent variational principle. The self-consistent single-particle hamiltonian $\hat{h}(t)$ is constructed from the Skyrme energy density functional. In this work, we employ the SLy5 Skyrme parametrization~\cite{chabanat1998a}. It is crucial to note that the standard Sky3D code~\cite{abhishek2024}  does not include the contributions from the squared spin-current tensor density (the so-called $J^2$  terms) that were included in the SLy5 fit. As a result, a fully consistent realization of the SLy5 energy density functional requires an explicit implementation of these $J^2$ terms. To ensure such consistency, we utilize a modified version of Sky3D in which the $J^2$ terms are explicitly incorporated, as has been successfully applied in previous works~\cite{dai2014a,shi2017,guo2018,guo2018b,sun2022,li2022,wu2022,sun2023,li2024c,huang2024b}. The set of coupled nonlinear TDHF equations is solved on a three-dimensional Cartesian coordinate system without any symmetry restrictions.

Within this coordinate representation, the initial state for the time evolution is constructed from the ground-state wave functions of the projectile and target nuclei, obtained from static Hartree-Fock (HF) calculations with the same SLy5 functional. Nuclear multipole deformations are thus naturally incorporated into the initial states. Table~\ref{tab:deformation} summarizes the corresponding deformation parameters. These static properties determine the initial geometric configurations used in the subsequent TDHF calculations. For the projectile ${}^{54}\mathrm{Cr}$, the HF ground state is prolately deformed, with $\beta_{20}\approx0.218$ and negligible triaxiality. Its orientation can therefore be specified by a single polar angle $\theta_{\mathrm P}$ between its symmetry axis and the incident direction. We consider the two limiting configurations: the tip orientation, $\theta_{\mathrm P}=0^\circ$, where the symmetry axis is parallel to the incident direction, and the side orientation, $\theta_{\mathrm P}=90^\circ$, where it is perpendicular to the incident direction. For the target ${}^{243}\mathrm{Am}$, the HF ground state is also axially symmetric, but reflection asymmetric. As listed in Table~\ref{tab:deformation}, it exhibits a large quadrupole deformation, $\beta_{20}\approx0.282$, a sizable hexadecapole component, $\beta_{40}\approx0.132$, and nonvanishing odd-multipole deformations $\beta_{30}$ and $\beta_{50}$, which give rise to a pear-like shape. Consequently, the configurations with $\theta_{\mathrm T}=0^\circ$ and $180^\circ$ are physically inequivalent, corresponding to the broad and narrow ends of the target facing the projectile, respectively. To account for this reflection asymmetry, the target orientation is sampled over the full range $0^\circ\le\theta_{\mathrm T}\le180^\circ$. In practice, we consider nine target orientations: $0^\circ$, $30^\circ$, $45^\circ$, $60^\circ$, $90^\circ$, $120^\circ$, $135^\circ$, $150^\circ$, and $180^\circ$.

In all cases, the incident direction is chosen as the $x$ axis, and the intrinsic symmetry axes of both nuclei are placed in the reaction plane, taken to be the $x$-$z$ plane. The orientation angles $\theta_{\mathrm P}$ and $\theta_{\mathrm T}$ are therefore the polar angles of the projectile and target symmetry axes with respect to the $x$ axis, while their azimuthal angles are fixed. The TDHF calculations are performed in a numerical box of $60\times 28\times 48$~fm$^{3}$, with a spatial grid spacing of 1.0~fm and a time step of 0.2~fm/c. The initial separation between the centers of mass of the projectile and target is set to 26~fm. The collision is initialized by applying a boost corresponding to the chosen center-of-mass energy $E_{\mathrm{c.m.}}$, assuming an initial Coulomb trajectory. For each specified orientation pair $(\theta_{\mathrm P},\theta_{\mathrm T})$, the impact parameter is taken in the reaction plane and scanned along the $z$ axis, i.e., perpendicular to the incident $x$ direction. The impact parameter is scanned from $b=0$~fm in steps of $\Delta b=0.5$~fm, so that the corresponding orbital angular momentum increases with $b$. In other words, for each orientation pair the present calculations consider only one in-plane direction of the impact parameter and do not include an additional azimuthal average over other possible directions or out-of-plane rotations. The scan is extended to progressively larger $b$ values until only quasielastic events remain, namely when the outgoing fragments stay close to the entrance-channel masses and charges and only small nucleon transfer is observed. Thus, the TDHF simulations with large $b$ probes the transition from dissipative quasifission dynamics to nearly elastic scattering. The time evolution is terminated when the relative distance between the outgoing fragments exceeds 26~fm, after which the asymptotic scattering angle is obtained from Coulomb-trajectory propagation.

\begin{table}[tbp]
	\caption{Ground-state deformation parameters for the projectile $^{54}\text{Cr}$ and the target $^{243}\text{Am}$, calculated using the Hartree-Fock method with the Skyrme SLy5 parametrization.}
	\label{tab:deformation}
	\begin{ruledtabular}
		\begin{tabular}{ccccccc}
			Nucleus & $\beta_{20}$& $\beta_{22}$& $\beta_{30}$ & $\beta_{40}$ & $\beta_{50}$ & $\beta_{60}$   \\
			\hline 	
			$^{54}$Cr    & 0.218 & 0.000 &  0.000 & 0.063 & 0.000  &  -0.004 	\rule[0pt]{0pt}{2.6ex} \\
			$^{243}$Am   & 0.282 & 0.000 & -0.053 & 0.132 & -0.023 & 0.022 \\	
		\end{tabular}
	\end{ruledtabular}
\end{table}

The angle-averaged yield $\sigma_{\lambda}$ for a specific reaction channel $\lambda$ is quantified by integrating over the impact parameter and averaging over all initial orientations,
\begin{equation}
	\sigma_{\lambda} \propto \int_{b_{\text{min}}}^{b_{\text{max}}} b \, \text{d} b \int_{0}^{\frac{\pi}{2}}\text{d}\theta_{\text{P}} \sin(\theta_{\text{P}}) \int_{0}^{\pi} \text{d}\theta_{\text{T}} \sin(\theta_{\text{T}}) P_{b}^{\lambda}(\theta_{\text{P}},\theta_{\text{T}}),
\end{equation}
where $P_{b}^{\lambda}(\theta_{\text{P}},\theta_{\text{T}})$ is the probability (either 0 or 1 for a single TDHF run) that the event at a given impact parameter $b$ and orientation configuration ($\theta_{\text{P}},\theta_{\text{T}}$) results in channel $\lambda$.

\section{RESULTS AND DISCUSSION}
\label{results}
\subsection{Orientation dependence}
\begin{figure}[ht!]
	\includegraphics[width=\columnwidth]{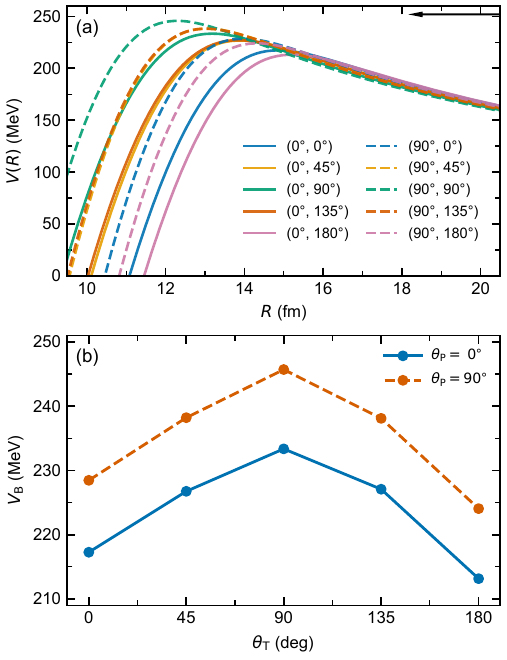}
	\caption{(a) Internuclear potentials for the $^{54}$Cr + $^{243}$Am reaction calculated within the FHF approximation for different orientation combinations $(\theta_{\mathrm{P}}, \theta_{\mathrm{T}})$. Solid and dashed lines correspond to $\theta_{\mathrm{P}} = 0^\circ$ and $90^\circ$, respectively, while curves of the same color share the same target orientation $\theta_{\mathrm{T}}$. The black arrow indicates the incident energy $E_{\mathrm{c.m.}} = 251.9$~MeV used in the dynamical simulations. (b) Barrier heights $V_B$ extracted from the FHF potentials as functions of the target angle $\theta_{\mathrm{T}}$ for the two projectile orientations $\theta_{\mathrm{P}} = 0^\circ$ and $90^\circ$.}
	\label{fig:fbcurve}
\end{figure}

\begin{figure}[t!]
	\includegraphics[width=\columnwidth]{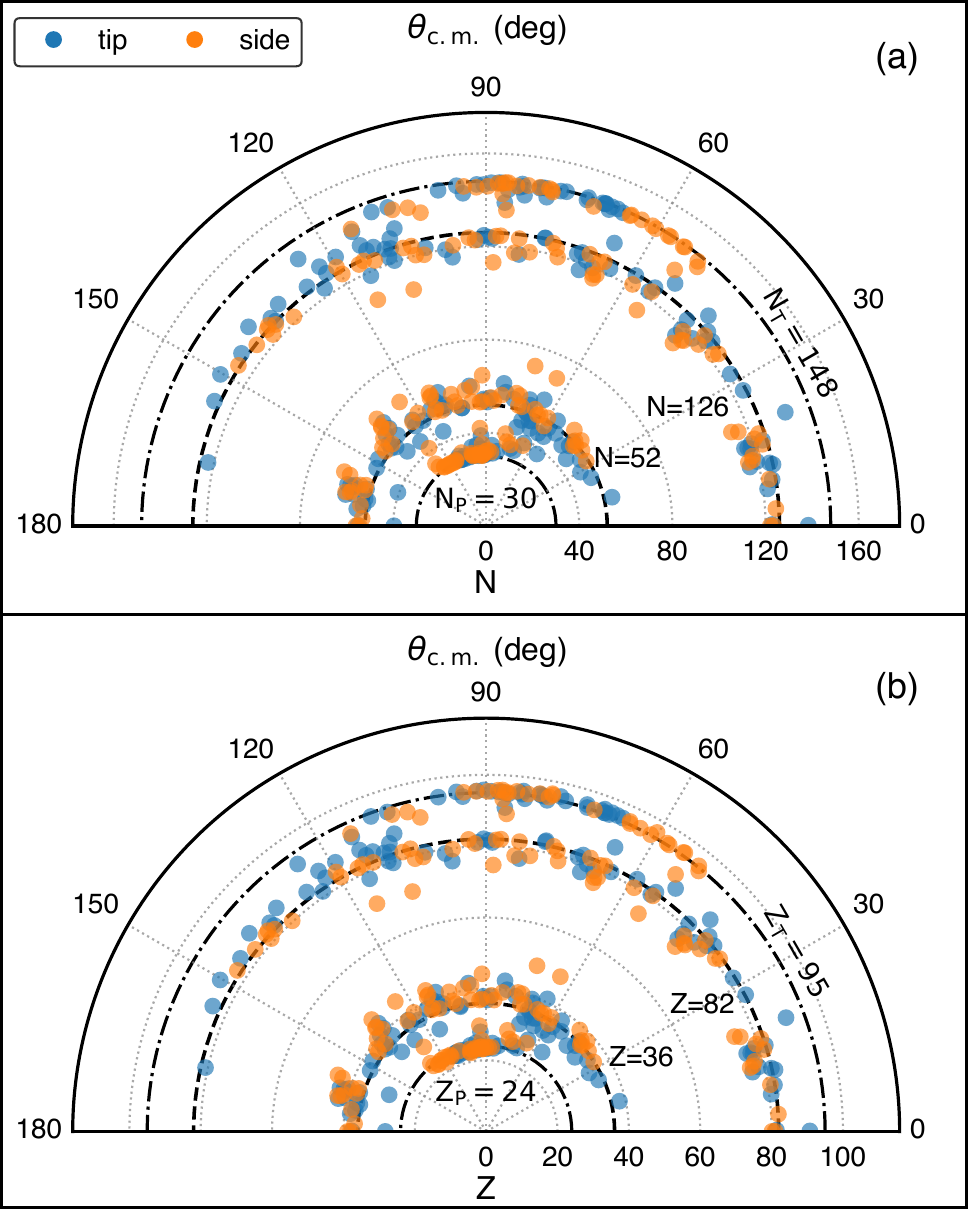}
	\caption{Polar representation of the final fragment distributions for the ${}^{54}\mathrm{Cr}+{}^{243}\mathrm{Am}$ reaction. The radial coordinate denotes (a) the neutron number $N$ and (b) the proton number $Z$, while the angular coordinate gives the center-of-mass scattering angle $\theta_{\mathrm{c.m.}}$ in degrees. Blue and orange symbols correspond to the tip and side orientations of the ${}^{54}\mathrm{Cr}$ projectile, respectively. The dashed circular arcs mark the positions of the relevant shell gaps, $N=52$, $126$ and $Z=36$, $82$, whereas the dash-dotted circular arcs labeled $N_{\mathrm{T}}$, $N_{\mathrm{P}}$, $Z_{\mathrm{T}}$, and $Z_{\mathrm{P}}$ indicate the initial neutron and proton numbers of the target and projectile.}
	\label{fig:polar}
\end{figure}
\begin{figure*}[ht!]
	\includegraphics[width=\textwidth]{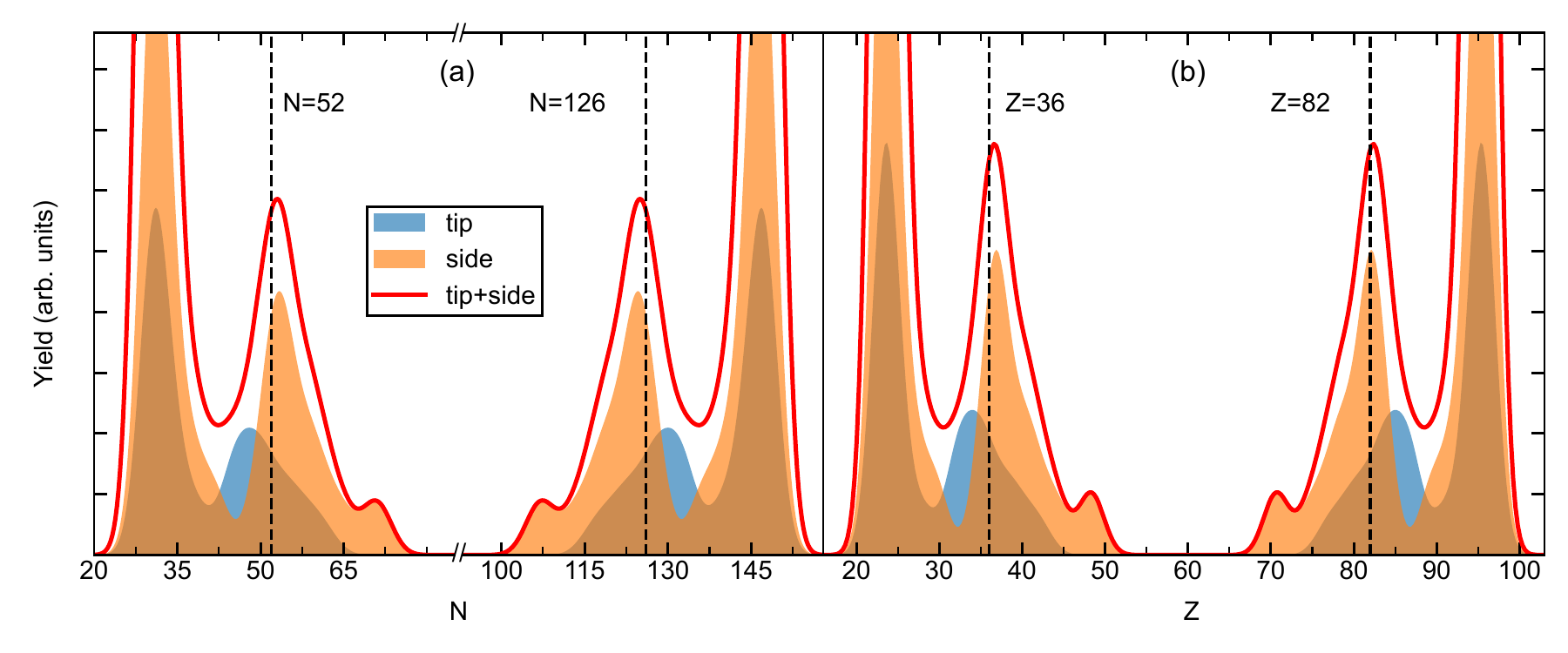}
	\caption{Fragment yield distributions for the $^{54}\mathrm{Cr}+{}^{243}\mathrm{Am}$ reaction at $E_{\text{c.m.}} = 251.9$ MeV. The panels display the yields as a function of (a) neutron number $N$ and (b) proton number $Z$. The shaded regions represent the weighted contributions from the ${}^{54}\text{Cr}$ tip (blue) and side (orange) orientations, integrated over all contributing impact parameters and nine ${}^{243}\text{Am}$ target orientations. The red solid lines denote the total angle-integrated yields. Vertical dashed lines mark the positions of relevant shell gaps at $N=52, 126$ and $Z=36, 82$.}
	\label{fig:yield}
\end{figure*}

To interpret the subsequent dynamical simulations, we first analyze the entrance-channel properties through the nucleus-nucleus interaction potentials $V(R)$. Figure~\ref{fig:fbcurve}(a) shows these internuclear potentials for the ${}^{54}\text{Cr}+{}^{243}\text{Am}$ system, calculated within the frozen Hartree-Fock (FHF) approximation for different orientation combinations~\cite{washiyama2008,guo2012,guo2018}, while Fig.~\ref{fig:fbcurve}(b) summarizes the corresponding barrier heights $V_B$ extracted from panel (a) as functions of the target angle $\theta_{\mathrm{T}}$ for $\theta_{\mathrm{P}}=0^\circ$ and $90^\circ$. 

Since the FHF approximation neglects dynamical rearrangement effects during the collision, such as density polarization, neck formation, and shape evolution, it is expected to somewhat overestimate the actual dynamical barriers. Around the barrier radius, the projectile-target overlap remains relatively small, so the incomplete treatment of inter-nuclear antisymmetrization effects in the FHF approximation is expected to have very limited influence on the barrier. These effects should become more important at shorter distances, where the density overlap is substantially large. A more dynamical treatment such as DC-TDHF would generally give lower effective barriers, but these are not strictly unique and may exhibit an incident-energy dependence~\cite{umar2010a,umar2014a}. We did DC-TDHF calculations for the representative orientation $(\theta_{\mathrm P},\theta_{\mathrm T})=(0^\circ,90^\circ)$, which give barrier heights of 228.52, 229.41, and 230.37 MeV at $E_{\mathrm{c.m.}}=240$, 245, and 250 MeV, respectively, whereas the corresponding FHF barrier for the same orientation is 233.23 MeV. Since the full DC-TDHF simulations for all orientations would not only be very time-consuming, but also require specifying the incident-energy protocol used to extract the dynamical barriers, we use FHF results as a common static baseline for comparing different initial orientations and for guiding the choice of the incident energy. The incident energy $E_{\mathrm{c.m.}}=251.9$ MeV, indicated by the arrow in Fig.~\ref{fig:fbcurve}(a), was thus chosen to lie slightly above the highest barrier obtained within the FHF approximation.

Figure~\ref{fig:fbcurve}(a) shows that the interaction barrier depends sensitively on the initial orientations, and Fig.~\ref{fig:fbcurve}(b) makes this trend more transparent. Overall, for a given target orientation $\theta_{\mathrm{T}}$, the barriers for $\theta_{\mathrm{P}}=90^\circ$ are systematically higher than those for $\theta_{\mathrm{P}}=0^\circ$, indicating that the projectile side orientation generally leads to stronger Coulomb hindrance. Accordingly, the highest barrier is obtained for the $(\theta_{\mathrm{P}},\theta_{\mathrm{T}})=(90^\circ,90^\circ)$ configuration, corresponding to the side-side geometry. In addition, for a fixed projectile orientation, the barrier heights for opposite target orientations are not identical: for both $\theta_{\mathrm{P}}=0^\circ$ and $90^\circ$, the barrier at $\theta_{\mathrm{T}}=0^\circ$ is systematically higher than that at $\theta_{\mathrm{T}}=180^\circ$. This asymmetry reflects the reflection-asymmetric shape of target nucleus ${}^{243}\mathrm{Am}$ associated with its non-negligible octupole deformation, while the difference becomes less pronounced at intermediate target angles. Such a pronounced orientation dependence is consistent with previous phenomenological studies emphasizing that the mutual orientation of deformed nuclei plays a key role in shaping the entrance-channel barrier and the hindrance of complete fusion~\cite{nasirov2026a}. The present microscopic calculation supports the same trend.

Following the static analysis in Fig.~\ref{fig:fbcurve}, we investigate the dynamical evolution using the TDHF theory. The calculated fragment scattering angle distributions are displayed in Fig.~\ref{fig:polar}. The radial coordinate corresponds to the neutron number $N$ and proton number $Z$ for panels (a) and (b), respectively. Each symbol represents a single TDHF trajectory at a given impact parameter and is plotted with equal visual weight. The inner and outer branches correspond to the projectile-like fragments (PLFs) and target-like fragments (TLFs), respectively, with blue and orange symbols denoting the tip and side orientations of the ${}^{54}\text{Cr}$ projectile. In Fig.~\ref{fig:polar}(a), the points lying close to the initial-value arcs correspond to quasielastic scattering events, forming narrow bands with nearly constant radial coordinates. Semicircular arcs at $N = 52$ and $N = 126$ are drawn to highlight the locations of the deformed and spherical shell closures, respectively. The target-like QF products are primarily distributed in the vicinity of the spherical $N = 126$ shell, spanning a broad angular range from $0^\circ$ to $150^\circ$. A closer inspection reveals that most TLF points are located along and slightly inside the $N = 126$ arc, while the corresponding PLF points are distributed just outside the deformed $N = 52$ arc. This alignment is consistent with previous studies highlighting the stabilizing role of the octupole-deformed shell gaps in the $N = 52$--$56$ region during fission and quasifission processes~\cite{scamps2019,simenel2021,godbey2019}. In Fig.~\ref{fig:polar}(b), the angular behavior of the proton number $Z$ mirrors the neutron distribution observed in panel (a). The QF events exhibit a significant drift in proton number away from the entrance-channel values $Z_{\text{P}}=24$ and $Z_{\text{T}}=95$, with the fragments accumulating along the circular arcs at $Z=36$ and $Z=82$, which mark the positions of the deformed and spherical shells, respectively~\cite{kozulin2022,morjean2017}. Collectively, these observations indicate that the QF fragments are steered by strong shell effects, associated with both spherical closures and deformed shell gaps.

Complementing the angular view in Fig.~\ref{fig:polar}, the integrated yields allow for a precise quantitative determination of the most probable nucleon numbers. Figure~\ref{fig:yield} presents these projected distributions for the neutron number $N$ in panel (a) and proton number $Z$ in panel (b), obtained by integrating over all contributing impact parameters and the nine sampled ${}^{243}\text{Am}$ target orientations. Note that the intense peaks at the extremities, corresponding to quasielastic events, are truncated to enhance the visibility of the QF components, which are the primary focus of this study.
We focus first on the projectile side configuration represented by the orange shaded area. In Fig.~\ref{fig:yield}(a), the heavy fragment yield peaks slightly below the $N = 126$ line. In the complementary region, the light fragment distribution lies predominantly above the $N = 52$ line, with its peak situated at $N = 54$. This pattern may suggest that the deformed shell gaps in the $N = 52\text{--}56$ region play a crucial role in anchoring the light fragment. In Fig.~\ref{fig:yield}(b), the heavy fragment yield is sharply centered at the spherical $Z = 82$ shell closure. Simultaneously, the peak for the light fragment is located slightly above the vertical dashed line at $Z = 36$.

While these observations might initially suggest a dominant role of the spherical $Z = 82$ shell in shaping the proton channel, disentangling its specific influence from that of the complementary light fragment presents a fundamental challenge. This ambiguity arises because the total nucleon numbers are rigorously conserved during the binary breakup process. As a result, the formation of a specific heavy fragment automatically dictates the composition of its complementary light partner, and vice versa. Consequently, one cannot definitively distinguish whether the mass equilibration is stopped primarily by the spherical closure of the heavy fragment, the deformed shell gaps of the light fragment, or a cooperative interplay between these coupled shell effects.

\begin{figure}[t!]
	\includegraphics[width=\columnwidth]{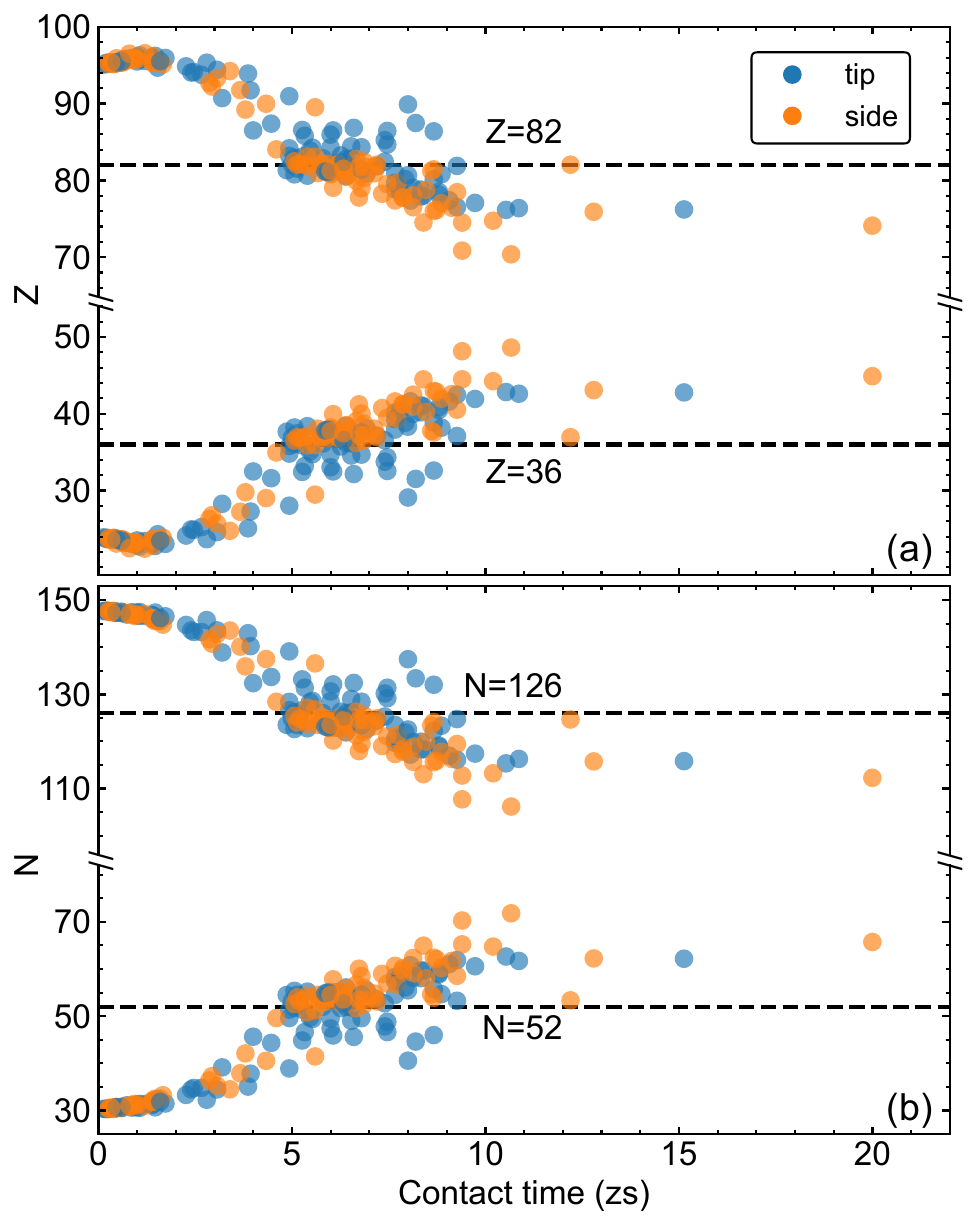}
	\caption{Fragment (a) proton number $Z$ and (b) neutron number $N$ as a function of contact time for the ${}^{54}\text{Cr}+{}^{243}\text{Am}$ reaction at $E_{\text{c.m.}} = 251.9$ MeV. The results for the projectile tip (blue points) and side (orange points) orientations are shown separately. Horizontal dashed lines indicate the positions of key spherical and deformed shell gaps.}
	\label{fig:cont}
\end{figure}
\begin{figure}[t!]
	\includegraphics[width=\columnwidth]{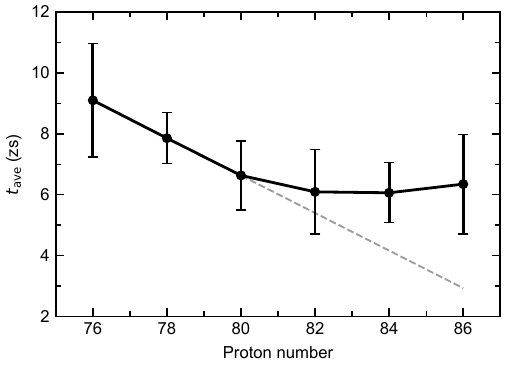}
	\caption{Average contact time $t_{\text{ave}}$ as a function of the heavy-fragment proton number $Z$ for the $^{54}\mathrm{Cr}+{}^{243}\mathrm{Am}$ reaction. The results are grouped into bins of width $\Delta Z = 2$ centered on even proton numbers. Only bins containing more than five events are shown. The solid points denote the average contact time, and the error bars correspond to the standard deviation. The dashed line serves as a guide to the eye.}
	\label{fig:ave}
\end{figure}

In sharp contrast to the strong shell-driven features observed in the side orientation, the tip orientation (blue shades) exhibits fundamentally different dynamics. As shown in Fig.~\ref{fig:yield}, while the side configuration is characterized by sharp, shell-stabilized peaks, the tip yield distribution does not exhibit these prominent features. Instead, the yields are dispersed in the intermediate region bounded by the vertical shell lines and the quasielastic peaks, indicating a lesser degree of net nucleon transfer. This trend is consistent with the observations in Fig.~\ref{fig:polar}, where the tip-orientation data points are broadly distributed between the shell closures and the quasielastic area. Since the absolute magnitude of these yields is further modulated by the geometrical weighting factor $\sin\theta_{\text{P}}$ inherent in the cross-section calculation, the contribution from tip collisions is significantly suppressed compared with that from side collisions. Consequently, the total yield distribution (red solid line), which represents the sum of the weighted contributions from all orientations, is overwhelmingly determined by the shell-driven dynamics of the side configuration. We note, however, that standard TDHF may be less accurate for channels involving large mass and charge transfer due to the lack of quantum fluctuation. Therefore, the present results are best viewed as illustrating the underlying trends and shell-driven mechanisms, rather than as fully quantitative predictions of the corresponding yields~\cite{sekizawa2013,sekizawa2019}.

The reaction dynamics are further elucidated in Figure~\ref{fig:cont}, which displays the final fragment proton number $Z$ in panel (a) and neutron number $N$ in panel (b) as a function of the contact time. The contact time is defined as the time interval during which the minimum density in the neck region exceeds 0.03~fm$^{-3}$. The data reveal that the QF process occurs on a rapid timescale, with the vast majority of QF events concentrated within a contact time window of 5 to 10~zs. Only a small fraction of events exhibit prolonged interaction times, extending up to approximately 20~zs. Comparing the projectile orientations, we observe that the overall duration of the interaction is similar for both the tip and side configurations. However, a notable difference emerges in the nucleon exchange dynamics. Specifically, for a given contact time, the side orientation (orange points) consistently exhibits a larger magnitude of nucleon transfer compared to the tip orientation (blue points). This observation implies that the side configuration, characterized by a more compact contact geometry, facilitates a faster rate of mass equilibration. Together with the fragment localization discussed in Fig.~\ref{fig:polar}, this trend supports a picture in which the quasifission evolution is guided by the competing shell-favored asymmetric valleys on the compound-nuclear potential-energy surface, rather than by a monotonic drift toward full mass equilibration~\cite{mcGlynn2023,lee2024}.

To extract the mean dynamical trends governing the nucleon transfer, Fig.~\ref{fig:ave} focuses on the statistical properties of the target-like fragments in the dominant region around $Z\approx 82$. Since the TDHF simulations provide non-integer expectation values for the final fragment proton number, rather than discrete integer particle numbers, the results are grouped into bins of width $\Delta Z = 2$ centered on even values of $Z$. To ensure sufficient statistics, only bins containing more than five events are retained in the analysis; in practice, this corresponds to the interval $Z=76$--$86$ shown in Fig.~\ref{fig:ave}. The black points represent the average contact time $t_{\text{ave}}$ for each bin, while the error bars denote the standard deviation, reflecting the spread of contact times among trajectories leading to similar final fragments.

In the range $Z=76$--$80$, $t_{\text{ave}}$ increases approximately linearly as the heavy fragment moves further away from the entrance channel, which is consistent with a classical diffusion picture in which a larger nucleon transfer requires a longer interaction time. A dashed line extending from $Z=80$ toward the $Z=86$ region is included to guide the eye. By contrast, as the proton number approaches the shell closure at $Z=82$ from above, the transferred nucleon number continues to increase, whereas the average contact time remains nearly constant or even decreases. This deviation from the classical diffusion trend indicates the influence of shell effects. As the heavy fragment approaches the magic number $Z=82$, the shell structure tends to hinder further shape evolution toward a more elongated dinuclear configuration, thereby promoting an earlier neck rupture and scission. Consequently, the $Z=82$--$86$ region exhibits a plateau-like behavior of $t_{\text{ave}}$ instead of the monotonic increase expected from a purely diffusive process.

\subsection{Energy dependence for the tip-tip orientation}
\begin{figure}[t!]
	\includegraphics[width=\columnwidth]{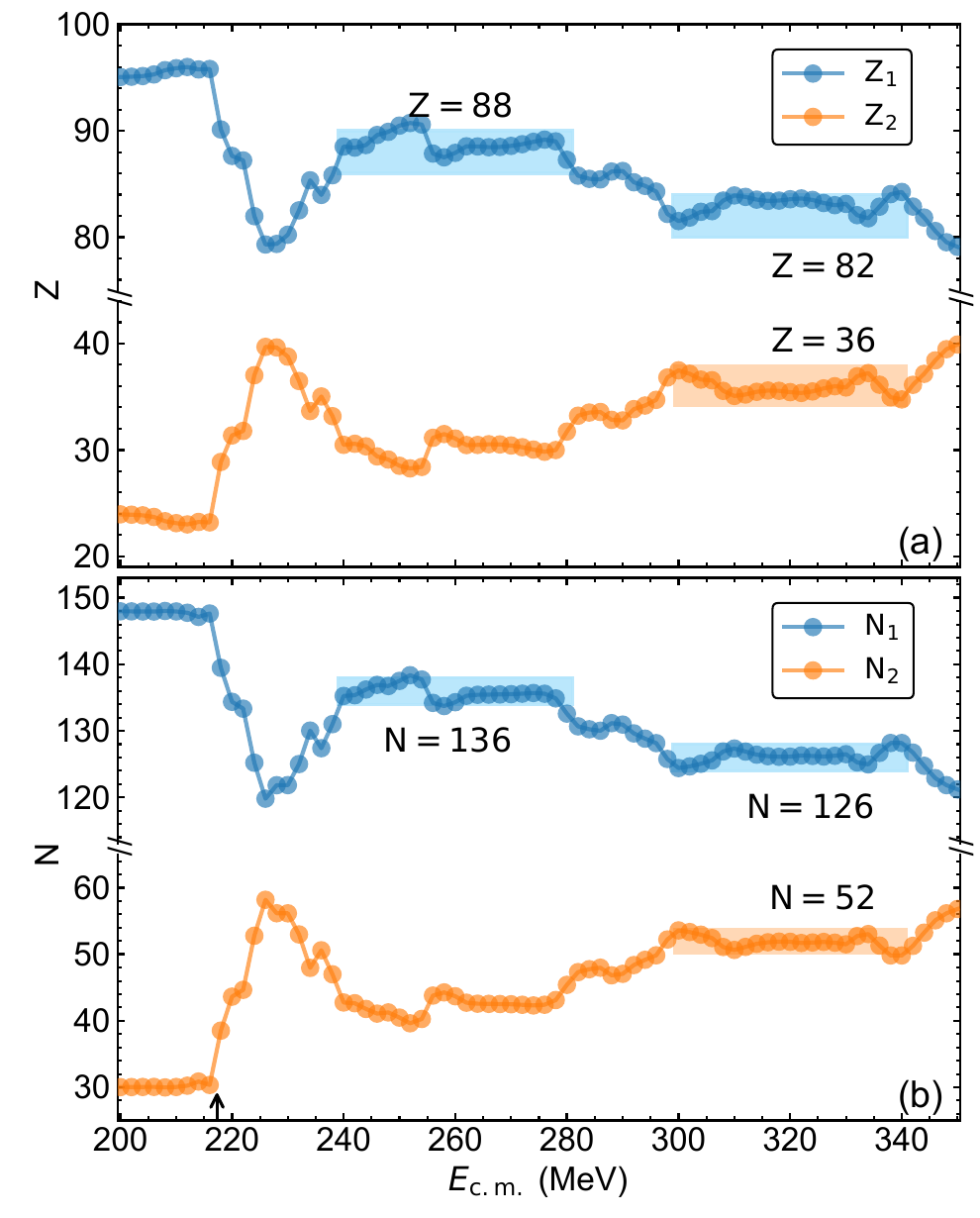}
	\caption{Fragment (a) proton number $Z_{1,2}$  and  (b) neutron number $N_{1,2}$  as a function of the incident $E_{\text{c.m.}}$ for the ${}^{54}\text{Cr}+{}^{243}\text{Am}$ reaction. The results are obtained for central collisions in the fixed tip-tip orientation. The horizontal shaded bands mark the specific shell regions discussed in the text, defined by a width of $\pm 2$ centered at $Z=88$ and $N=136$ for the octupole-deformed region, $Z=82$ and $N=126$ for the spherical shell closure, and $Z=36$ and $N=52$ for the relevant deformed shells of the light partner. The black arrow indicates the position of the Coulomb barrier at $V_{\text{B}} = 217.3$~MeV calculated using the FHF approach.}
	\label{fig:ecm}
\end{figure}

Having analyzed the orientation dependence at a fixed energy, we now turn to the incident-energy dependence of the QF dynamics over a broad range from below to well above the Coulomb barrier. In this part, we restrict ourselves to the tip-tip configuration. This choice does not imply that the tip-tip channel dominates the orientation-averaged QF yield; indeed, at $E_{\mathrm{c.m.}}=251.9$ MeV the side configuration exhibits stronger transfer and contributes more significantly to the total yield. The reason for focusing on the tip-tip channel is instead that, for near-central collisions, it exhibits a qualitatively different behavior from the other orientations, namely a smaller nucleon transfer. This reduced transfer is consistent with the tendency of the heavy fragment  localized around $Z\approx 88$ and $N\approx 136$, possibly owing to the influence of octupole-deformed shell effects in this region. The tip-tip configuration therefore provides a useful channel for tracking how this distinct shell-guided reaction path evolves with incident energy.

The simulations cover the range $E_{\mathrm{c.m.}}=200$--$350$ MeV with a fine grid step of $2$ MeV. As shown in Fig.~\ref{fig:ecm}, the reaction is dominated by quasielastic scattering below the barrier. Once the energy rises above the tip-tip barrier, the nucleon transfer increases rapidly in the near-barrier region around $218$--$226$ MeV. As the energy increases further, the transferred nucleon number decreases in the interval $226$--$240$ MeV and then enters a broad plateau from about $240$ to $280$ MeV, with the heavy fragment stabilizing within the blue shaded regions centered at $Z=88$ and $N=136$. Notably, these nucleon numbers correspond to $^{224}\text{Ra}$, which lies within the well-known region of octupole deformation~\cite{butler2020, wollersheim1993}. This suggests that the reaction product is anchored by the stabilizing effects of octupole shell gaps in this energy window. At higher energies from $280$ to $300$~MeV, the nucleon transfer rises again and eventually enters a regime above $300$~MeV where the reaction products exhibit distinct shell features. In this region, the heavy fragments fall within the shaded bands corresponding to the spherical shell closures at $Z=82$ and $N=126$. Simultaneously, the complementary light fragments are found within the deformed shell ranges around $Z=36$ and $N=52$. These observations suggest that the dynamics in this energy domain are likely shaped by the concurrent influence of spherical closures in the heavy fragment and deformed shell gaps in the light partner.

The clear shift from the octupole-deformed configuration around $Z \approx 88$ and $N \approx 136$ observed at intermediate energies to the spherical configuration around $Z \approx 82$ and $N \approx 126$ at higher energies explicitly illustrates the energy dependence of shell effects. This transition parallels the shape evolution with temperature predicted in nuclear structure studies~\cite{zhang2017}, suggesting that the octupole shell effects are more vulnerable to thermal damping and diminish at higher excitation energies, thereby allowing the reaction to be steered by the more resilient spherical shell closure. Collectively, these findings reveal that the reaction is governed not only by the interplay between the spherical and deformed shells of the complementary partners but also by the evolution of the dominant shell structure with incident energy. Understanding this dependence is important for SHE synthesis, as QF is the primary hindrance to fusion. Our results suggest that, by carefully selecting the incident energy, one may reduce the tendency toward strongly shell-driven QF. For instance, within the present tip-tip analysis with the SLy5 functional, the region around $E_{\text{c.m.}} \approx 226$~MeV emerges as an interesting candidate region for reduced shell-driven QF, since the steering influence of shell effects appears to be relatively weakened there, while the corresponding excitation energy remains relatively low, which may be more favorable for the survival of the compound nucleus against fission.

However, the present energy-window feature should be interpreted as a channel-specific observation rather than a definitive optimization of the beam energy. In particular, it has been identified only for the tip-tip orientation, whereas the side-related orientations contribute more to the total yield. A systematic energy-dependent study of projectile-side orientations will therefore be needed to establish whether a similar feature persists in the  side-dominated channels. We also note that the present observation has been obtained with the SLy5 functional, and its robustness with respect to the choice of energy density functional remains to be examined in future work.

\section{SUMMARY AND CONCLUSIONS}
\label{summary}
In this work, we have performed a comprehensive microscopic investigation of QF dynamics in the ${}^{54}\text{Cr}+{}^{243}\text{Am}$ reaction, a key candidate for synthesizing SHE 119, using the TDHF theory. We systematically explored the influence of projectile-target orientations and the role of incident energy on the reaction outcomes. Our static FHF calculations, used here as a common reference for the entrance channel,  show that the entrance-channel potential is highly sensitive to the initial orientations, particularly due to the octupole deformation of the ${}^{243}\text{Am}$ target, which breaks reflection symmetry.  The full TDHF dynamical simulations reveal a pronounced dependence of the QF mechanism on the projectile orientation. At the reference energy, for side collisions of ${}^{54}\text{Cr}$, which are found to dominate the total QF yield, the fragment distributions are strongly governed by shell effects. The heavy fragment production is concentrated near the $Z=82$ spherical shell closure, while the light fragment production is driven by the $N=52 \text{--} 56$ deformed shell. In contrast, tip collisions result in relatively smaller nucleon transfer and do not display prominent signatures of shell effects or their competition in the fragment yield distributions. A statistical analysis of the contact times shows that, as the heavy fragment proton number approaches the $Z=82$ magic shell, the average contact time remains  nearly  constant or even decreases, contrary to the classical diffusion expectation that more nucleon transfer requires longer time. This behavior is attributed to the quantum shell effect, which provides stability to the fragment's shape, making it less prone to deformation, thereby accelerating neck rupture and reducing the contact time. Furthermore, our investigation of the energy dependence for tip-tip central collisions reveals a complex relationship between the incident energy and shell effects in QF. In different energy regions, the reaction dynamics are governed by distinct shell modes, including the octupole-deformed region around $Z\approx 88$ and $N\approx 136$ at intermediate energies and the spherical-shell region around $Z\approx 82$ and $N\approx 126$ at higher energies. This behavior demonstrates that the dominant shell structure guiding the QF path may change with incident energy. Such energy dependence is relevant for SHE synthesis, since the near-barrier region may simultaneously weaken shell-driven QF and correspond to relatively low excitation energies, which are more favorable for the survival of the compound nucleus against fission.

In summary, our microscopic study reveals that the manifestation of shell effects in the final QF fragments is a dynamical outcome that depends sensitively on both the initial orientation and the incident energy. The present energy-window feature has been identified within the tip-tip analysis using the SLy5 functional, and its generality with respect to other orientations and energy density functionals remains to be clarified.

\begin{acknowledgments}
This work has been supported by the Strategic Priority Research Program of the Chinese Academy of Sciences (Grant No. XDB1550100), and the National Natural Science Foundation of China (Grants No.12435008, No.12375127, and No.12205308).
\end{acknowledgments}

\section*{DATA AVAILABILITY}
The data that support the findings of this article are not publicly available. The data are available from the authors upon reasonable request.

\bibliography{QF119.bib}
\end{document}